\documentclass[
 reprint,
 superscriptaddress,
 amsmath,amssymb,
 aps,
 pra,
]{revtex4-2}

\usepackage{graphicx}
\usepackage{dcolumn}
\usepackage{bm}
\usepackage{hyperref}
\usepackage{siunitx}
\usepackage{xspace}
\usepackage{xcolor}

\newcommand{\ket}[1]{\ensuremath{\lvert #1 \rangle}\xspace}%
\newcommand{\bra}[1]{\ensuremath{\langle #1 \rvert}\xspace}%
\newcommand{\braket}[2]{\ensuremath{\langle #1 \rvert #2\rangle}\xspace}%

\begin{document}

\title{Realization of a Rydberg-dressed extended Bose Hubbard model}

\author{Pascal~Weckesser}
    \affiliation{Max-Planck-Institut f\"{u}r Quantenoptik, 85748 Garching, Germany}
    \affiliation{Munich Center for Quantum Science and Technology (MCQST), 80799 Munich, Germany}
\author{Kritsana Srakaew}
    \affiliation{Max-Planck-Institut f\"{u}r Quantenoptik, 85748 Garching, Germany}
    \affiliation{Munich Center for Quantum Science and Technology (MCQST), 80799 Munich, Germany}
\author{Tizian Blatz}
    \affiliation{Munich Center for Quantum Science and Technology (MCQST), 80799 Munich, Germany}
    \affiliation{Department of Physics and Arnold Sommerfeld Center for Theoretical Physics (ASC), Ludwig-Maximilians-Universit\"{a}t M\"{u}nchen, 80333 M\"{u}nchen, Germany}
\author{David Wei}
    \affiliation{Max-Planck-Institut f\"{u}r Quantenoptik, 85748 Garching, Germany}
    \affiliation{Munich Center for Quantum Science and Technology (MCQST), 80799 Munich, Germany}
\author{Daniel Adler}
    \affiliation{Max-Planck-Institut f\"{u}r Quantenoptik, 85748 Garching, Germany}
    \affiliation{Munich Center for Quantum Science and Technology (MCQST), 80799 Munich, Germany}
\author{Suchita Agrawal}
    \affiliation{Max-Planck-Institut f\"{u}r Quantenoptik, 85748 Garching, Germany}
    \affiliation{Munich Center for Quantum Science and Technology (MCQST), 80799 Munich, Germany}
\author{Annabelle Bohrdt}
    \affiliation{Munich Center for Quantum Science and Technology (MCQST), 80799 Munich, Germany}
    \affiliation{Institute of Theoretical Physics, University of Regensburg, 93053 Regensburg, Germany}
\author{Immanuel Bloch}
    \affiliation{Max-Planck-Institut f\"{u}r Quantenoptik, 85748 Garching, Germany}
    \affiliation{Munich Center for Quantum Science and Technology (MCQST), 80799 Munich, Germany}
    \affiliation{Fakult\"{a}t f\"{u}r Physik, Ludwig-Maximilians-Universit\"{a}t, 80799 Munich, Germany}
\author{Johannes Zeiher}
	\affiliation{Max-Planck-Institut f\"{u}r Quantenoptik, 85748 Garching, Germany}
    \affiliation{Munich Center for Quantum Science and Technology (MCQST), 80799 Munich, Germany}
    \affiliation{Fakult\"{a}t f\"{u}r Physik, Ludwig-Maximilians-Universit\"{a}t, 80799 Munich, Germany}
\date{\today}

\begin{abstract}
    The competition of different length scales in quantum many-body systems leads to various novel phenomena, including the emergence of correlated dynamics or non-local order.
    To access and investigate such effects in an itinerant lattice-based quantum simulator, it has been proposed to introduce tunable extended-range interactions using off-resonant optical coupling to Rydberg states.
    %
    %
    However, experimental realizations of such ``Rydberg dressing" have so far mostly concentrated on spin systems without motion.
    Here, we overcome a number of experimental challenges limiting previous work and realize an effective one-dimensional extended Bose-Hubbard model (eBHM).
    Harnessing our quantum gas microscope, we probe the correlated out-of-equilibrium dynamics of extended-range repulsively-bound pairs at low filling, and kinetically-constrained ``hard rods" at half filling.
    Near equilibrium, we observe density ordering when adiabatically turning on the extended-range interactions.
    Our results demonstrate the versatility of Rydberg dressing in engineering itinerant optical lattice-based quantum simulators and pave the way to realizing novel light-controlled extended-range interacting quantum many-body systems.
\end{abstract}

\maketitle


%
Extended-range interactions between itinerant quantum particles give rise to a wealth of interesting phenomena in many-body systems, including, for example, the exotic properties of superfluid helium~\cite{Boninsegni2012}.
%
%
When trapped in lattices, such systems with interactions that span at least the next-nearest neighbor site are described by \emph{extended Hubbard models} (eHM), which feature a rich phenomenology arising from the extended-range interactions competing with the tunneling and local interaction energy scales~\cite{Defenu2023}.
Near equilibrium, manifestations of this competition include the emergence of non-local order~\cite{berg2008}, lattice supersolids~\cite{Cinti2014,Geissler2017}, and exotic cluster Luttinger liquid phases~\cite{Mattioli2013,Dalmonte2015}, while out-of-equilibrium the emergence of repulsively-bound pairs has been predicted~\cite{valiente2009}.
Although ultra-cold atoms are an ideal platform to realize Hubbard systems~\cite{bloch2008,gross2017}, the vast majority of work up to this point has focused on systems with on-site interactions~\cite{chin2010}.
Recently, much effort has been directed at overcoming this limitation and opening the stage for quantum simulations including long-range interactions.
Experimentally, long-range interacting quantum simulators for itinerant particles have been realized using electric dipole interactions in molecules~\cite{yan2013,moses2017,bigagli2023,rosenberg2022,christakis2023,carroll2024}, magnetic dipole interactions in lanthanide atoms~\cite{baier2016,kadau2016,chomaz2022,su2023}, and cavity-assisted all-to-all interactions~\cite{landig2016,leonard2017,helson2023}.
Furthermore, Rydberg quantum simulators, usually tailored to simulating spin models~\cite{Browaeys2020} or quantum computing~\cite{Saffman2016,Morgado2021}, have been used to simulate itinerant hard-core interacting bosons through a spin-boson mapping~\cite{deleseleuc2019}.
However, despite several spectacular advances, the microscopic control and readout of long-range interacting gases of itinerant particles has been limited, except for recently realized quantum gas microscopes for dilute molecular samples~\cite{christakis2023} and the observation of quantum solids in optical lattices with a tunable lattice constant~\cite{su2023}.
Harnessing the full set of capabilities of quantum gas microscopes, which include both microscopic detection \textit{and} single-particle control~\cite{gross2021a}, has remained an outstanding experimental challenge in eHM.

\begin{figure}[t!]
    \centering
    \includegraphics[width=0.5\textwidth]{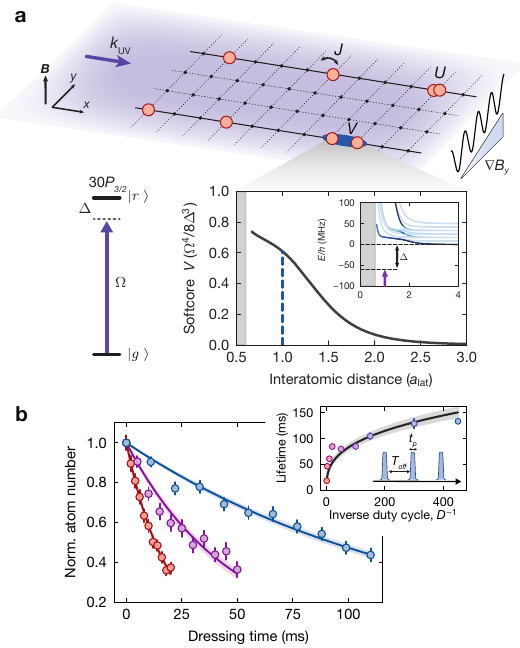}
    \caption{\textbf{Rydberg-dressed extended Bose Hubbard model and Rydberg lifetimes.}
    (\textbf{a}) Tilted lattice system and relevant energy scales.
    The atoms are allowed to tunnel along the chains with tunneling energy $J$ and experience repulsive on-site interactions $U$.
    We engineer long-range, nearest-neighbor interactions $V$ using off-resonant coupling to a Rydberg state $\ket{r}$, with Rabi frequency $\Omega = 2\pi\times \SI{20}{\mega\hertz}$ and detuning $\Delta = -2\pi\times\si{\numrange[range-phrase = -]{60}{400}}\,\si{\mega\hertz}$.
    The resulting softcore-shaped interaction can be derived by adding the contributions of various interaction potentials (see inset).
    Their relative optical coupling strength is indicated by the intensity of the blue coloring.
    (\textbf{b}) Duty-cycle dependent atomic lifetime for fixed $V/h = \SI{114(4)}{\hertz}$ in a unity-filled 1D chain of $11$ atoms. 
    Here we apply stroboscopic, pulsed dressing with pulse durations of $t_p = \si{\numrange[range-phrase = -]{0.5}{1}}\,\si{\micro\second}$ and tunable off-times $T_{\mathrm{off}} = \si{\numrange[range-phrase = -]{150}{300}}\,\si{\micro\second}$ (see inset).
    Scanning the duty cycle $D = t_p/(t_p+T_{\mathrm{off}})$ ranging from $D=1$ (red), over $D=1/3$ (purple), to $D=1/450$ (blue), we improve the atom lifetime by a factor of $\sim 7.2(3)$, following the expected scaling $\sqrt[3]{D^{-1}}$ (see inset).
    Averaged over all measurements, our experimental lifetimes reach $\SI{54(3)}{\percent}$ of their respective theoretical value.
    All error bars denote the s.e.m.
    }
    \label{fig:1}
\end{figure}
Off-resonant admixture of strongly interacting Rydberg states, or ``Rydberg dressing", offers an alternative pathway to realize extended-range interacting quantum gases that are compatible with in-situ control over single atoms and provides flexible interaction control.
Rydberg-dressed interactions emerge through the off-resonant coupling of short-range interacting ground state atoms to a higher-lying, long-range interacting Rydberg state with Rabi frequency $\Omega$ and detuning $\Delta $~\cite{Bouchoule2002, Henkel2010,Pupillo2010}.
Here, the Rydberg blockade modifies the AC Stark shift experienced by two nearby ground state atoms, which leads to a characteristic softcore-shaped potential with softcore height $V_0 = \Omega^4/(8 \Delta^3)$ and tunable interaction range $R_c = (\lvert C_6 \lvert / (2\hbar \lvert \Delta \lvert))^{1/6}$ of up to a few lattice sites, with an asymptotic van-der-Waals tail and coefficient $C_6$  (see Fig.~\ref{fig:1}a).
The relatively large interaction range $R_c$ easily spans micron-scale distances, making Rydberg dressing straightforwardly compatible with single-atom control \textit{and} detection in quantum gas microscopes~\cite{Mattioli2013,macri2014,vanbijnen2015,zhou2020,malz2023}, without the need to dynamically adjust the lattice spacing~\cite{su2023}.
The optical coupling also provides a unique handle to control the interaction, for example, to realize sudden quenches of the interaction strength on sub-microsecond timescales, periodically-modulated interactions~\cite{rapp2012}, or turning on the long-range interactions only after the system has been prepared.
Furthermore, employing blue- or red-detuned coupling enables the realization of either attractive or repulsive interactions~\cite{balewski2014}, and changing the external field and polarization of the coupling light allows for controlling the isotropy of the interaction~\cite{zeiher2016,steinert2023}.
So far, Rydberg-dressed interactions have been successfully implemented in different regimes for pinned atoms in the absence of motional dynamics to realize spin models~\cite{jau2016,zeiher2016,zeiher2017,hollerith2022,schine2022,steinert2023,eckner2023}. 
While first steps towards utilizing Rydberg-dressed interactions to realize itinerant fermionic systems were taken~\cite{guardado-sanchez2021}, a wide-spread adoption of the technique has been hindered by the observation of strong collective losses~\cite{goldschmidt2016,zeiher2016}.
These can be attributed to off-resonant scattering which can lead to the population of opposite-parity Rydberg states, whose interaction with the dressed state tunes the coupling light into resonance that can trigger an excitation avalanche~\cite{goldschmidt2016,zeiher2016,festa2022}.
Thus, up to this point, the feasibility of using Rydberg dressing to experimentally study such systems for tens of tunneling times or near equilibrium has been an open question.
In this work, we realize an effective one-dimensional extended Bose-Hubbard model (eBHM) using Rydberg-dressed $^{87}\mathrm{Rb}$ atoms in an optical lattice.
These results are enabled by
a stroboscopic dressing sequence that leads to significantly increased lifetimes of over $\SI{100}{\milli\second}$ in our system, surpassing previous observations by two orders of magnitude~\cite{zeiher2017,guardado-sanchez2021}.
We use the excellent optical control over the extended interaction strength in combination with microscopic control over single atoms to study the eBHM in a number of interesting regimes in- and out-of-equilibrium.
In particular, our studies explore several regimes of interaction strength and thus highlight the impact of this new energy scale on the emerging many-body physics. 
First, we observe the existence and dynamics of repulsively-bound pairs of atoms occupying neighboring lattice sites stabilized by strong extended-range interactions.
Second, increasing the filling, at large nearest-neighbor interaction, we realize a gas of ``hard rods" and demonstrate the striking difference in the density relaxation of an initial charge-density wave of such hard rods compared with the case of hard-core bosons.
Finally, we probe the nature of the low-energy states of the eBHM by slowly increasing the nearest-neighbor interaction and observing the emergence of density-ordering into a state similar to the quantum solids that were recently observed in magnetic dipolar atoms~\cite{su2023}.

\begin{figure*}[t!]
    \centering
    \includegraphics[width=\textwidth]{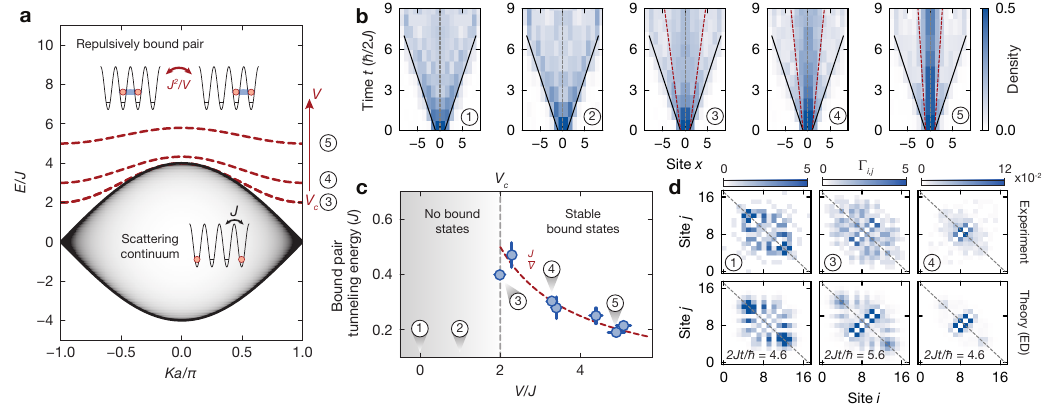}
    \caption{\textbf{Repulsively-bound pair states in the extended Bose-Hubbard model.}
    (\textbf{a}) Two-particle energy spectrum of the 1D eBHM as a function of center-of-mass quasi-momentum $K$.
    For $V \geq V_c = 2J$, the spectrum features stable bound states, illustrated as red dashed lines, with reduced tunneling energy $J^2/V$.
    (\textbf{b}) Symmetrized two-particle density dynamics for $V/J = \{0, 1.0(1) ,2.0(2), 3.3(1), 
    4.9(2)\}$ (cases \textcircled{1}-\textcircled{5}, respectively) post-selected on two particles being present.
    For $V < V_c$, the particles spread independently, following a ballistic light-cone with group velocity $2Ja_{\mathrm{lat}}/\hbar$ until the system's boundary is reached (see black line).
    For $V \geq V_c$, we observe the additional emergence of an inner, slower light-cone.
    The red dashed lines represent the best-fit result of the bound pair's expansion.
    (\textbf{c}) Extracted tunneling energy of the bound pair for variable $V$ in units of the single-particle tunneling energy $J$.
    We find excellent agreement with the exact solution $J^2/V$ derived in~\cite{valiente2009}, represented as the red dashed line.
    (\textbf{d}) Symmetrized two-particle correlator $\Gamma_{i,j}$ for variable $V$ and time.
    The upper (lower) row shows the measured correlator (the correlator calculated with ED).
    Without Rydberg-dressed interactions (case \textcircled{1}), we observe strong antibunching on the off-diagonal due to the ``fermionization" of the hard-core bosons.
    For $V \geq V_c$, we observe strong nearest-neighbor correlators on the diagonal, indicating correlated pair tunneling in good agreement with theory.
    The gray dashed line in (\textbf{b}) and (\textbf{d}) show the symmetrization axis.
    All error bars denote the s.e.m.
    }
    \label{fig:2}
\end{figure*}

\subsection*{Experimental system}

In our experiment, we prepare up to $200$ $^{87}\mathrm{Rb}$ atoms in the $\ket{g} =\ket{F = 2, m_F = +2}$ ground state in a single plane of a vertical optical lattice.
We realize a two-dimensional Bose-Hubbard model using a folded lattice~\cite{wei2023} with lattice spacing $a_\mathrm{lat} = \SI{752}{\nano\meter}$, tunable tunnel coupling $J$ and approximately fixed interaction strength $U/h = \SI{225}{\hertz}$ set by the scattering properties of the $^{87}\mathrm{Rb}$ atoms.
Using single-site addressing~\cite{weitenberg2011, fukuhara2013}, we prepare up to three parallel and independent one-dimensional (1D) systems oriented along the $x$-direction and separated by at least three lattice sites in the $y$-direction.
Each 1D system contains a variable and controllable number of atoms.
Tunneling along the $x$-direction is enabled at a rate $J$, while tunneling along the transverse $y$-direction is inhibited by applying a magnetic field gradient $\nabla B_y \simeq 350(3) \si{\hertz}/a_\mathrm{lat}$, see Fig.~\ref{fig:1}a.
In the following, all discussions refer to these equivalent 1D-systems.
We induce extended-range interactions up to two-lattice sites (see inset of Fig.~\ref{fig:1}a) by off-resonantly coupling atoms from the ground state $\ket{g}$ to the $\ket{r} = \ket{30\mathrm{P}_{\mathrm{3/2}};m_{j} = +3/2}$ Rydberg state with an ultraviolet laser at a wavelength of $\SI{297}{\nano\meter}$, which propagates along the $x$-direction.
We calibrate the resulting time-averaged interaction strength $V$ using a spectroscopic spin-echo sequence~\cite{SI}.
Our system is described by the Hamiltonian
\begin{equation}
         \hat{H} = 
          -J \sum_{\langle i,j \rangle} \hat{a}_i^{\dagger}\hat{a}_j +  \frac{U}{2} \sum_{i} \hat{n}_{i} (\hat{n}_{i} - 1) 
          +  V \sum_{i} \hat{n}_i \hat{n}_{i+1},
         \label{eq:eBHM}    
\end{equation}
where $\hat{a}_i^{\dagger}$ and $\hat{a}_i$ are the raising and lowering operator and $\hat{n}_i$ the number operator on site $i$.
In the present work, we focus on the 1D hard-core boson limit $U/J \gtrsim 11$, where doublon formation is strongly suppressed.
To maximize the advantages of controlling the extended-range interaction $V$ via the coupling light, we apply a stroboscopic pulse sequence with duty cycle $D$.
Compared with a continuous drive, our pulsed scheme enables us to operate at smaller detuning without suffering from detrimental collective losses.
The pulsed scheme improves the product of the time-averaged interaction times lifetime, as $V\tau\propto 1/\Delta$~\cite{Zeiher2017a, hines2023}, effectively scaling as $\sqrt[3]{D^{-1}}$ if the time-averaged softcore height is kept constant~\cite{SI}.
%
%
Our measurements confirm the drastically increased lifetime that extends beyond $\SI{100}{\milli\second}$ for $D \lesssim 1/150$, in agreement with the expected scaling (see inset of Fig.~\ref{fig:1}b).
Note that such enhanced lifetimes exceed our typical tunneling times $\hbar/(2J) \simeq \si{\numrange[range-phrase = -]{4}{7}}\,\si{\milli\second}$ by at least an order of magnitude while reaching $V/J \simeq \si{\numrange[range-phrase = -]{6}{10}}$, accomplishing a long-pursued milestone.
Surpassing previously reported lifetimes by about two orders of magnitude~\cite{zeiher2017,guardado-sanchez2021}, puts us in place to probe microscopic features of the eBHM experimentally.
%


\subsection*{Repulsively-bound pair states}
We first focus on out-of-equilibrium dynamics by preparing two atoms on neighboring sites in the 1D chain and quenching the tunneling energy to finite values.
In the eBHM, the interaction between atoms on adjacent sites has been predicted to result in the emergence of bound states when $V$ exceeds the critical value $V_c=2J$~\cite{valiente2009}.
These bound states arise through the interplay of strong nearest-neighbour interactions and the bounded spectrum of extended states in the optical lattice, which have been analogously observed for atoms bound by the on-site Hubbard interaction before~\cite{winkler2006,preiss2015}.
In the center-of-mass frame, the energy of the bound-state solution is given by $E_{B}(K) = V + 4J^2/V \mathrm{cos}(K a_{\mathrm{lat}}/2)$, with $K$ the center-of-mass quasimomentum~\cite{valiente2009}.
The bound state can thus be interpreted as an isolated state separated from the continuum of free single-particle states by the energy $V$ and featuring a bandwidth set by the reduced tunneling energy $J^2/V$.
This reduced tunneling energy of the bound pair can be intuitively understood as a hopping of the bound state occuring perturbatively in $J/V$ through a second-order process coupling off-resonantly to the continuum of free states.
In our experiment, we detect the presence of the bound state and measure its group velocity by monitoring the time evolution of an atom pair after a sudden quench of the tunnel coupling in the lattice (see Fig.~\ref{fig:2}b).
In the absence of Rydberg-dressed interactions ($V=0$; case \textcircled{1}), we observe a characteristic light-cone indicating the free expansion of the two particles expected for the hard-core boson limit~\cite{karski2009,sansoni2012,preiss2015}.
The expansion speed of the wavefront is consistent with the Lieb-Robinson bound given by $2Ja_{\mathrm{lat}}/\hbar$ (see black line in Fig.~\ref{fig:2}b).
When introducing nearest-neighbour interactions $V\geq V_c$~\cite{valiente2009}, we observe the emergence of a second light-cone with a lower expansion speed of the wavefront (cases \textcircled{3}-\textcircled{5}).
We determine the group velocities for the free and bound state by fitting an incoherent sum of four Bessel functions of the first kind to our experimental data~\cite{SI} and find excellent agreement with the expected bound pair tunneling energy $J^2/V$, see Fig.~\ref{fig:2}c).
To unequivocally prove the existence of the repulsively-bound pairs also microscopically, we evaluate the two-particle correlator $\Gamma_{i,j} = \langle a_i^{\dagger} a_j^{\dagger} a_i a_j  \rangle$~\cite{preiss2015}.
In absence of long-range interactions, the hard-core bosons effectively behave as noninteracting spinless fermions and show long-range correlations on the anti-diagonal.
The latter is a direct consequence of Hanbury Brown and Twiss interference for fermions, resulting in pronounced anti-bunching~\cite{preiss2015}.
With nearest-neighbor interactions $V \geq V_c$, we observe correlated pair-tunneling, resulting in strong nearest-neighbour correlations on the diagonal.
Here, we find an overall good agreement with exact diagonalization (ED).


\subsection*{Constrained ``hard rod" dynamics}

\begin{figure}[t!]
    \centering
    \includegraphics[width= 0.5\textwidth]{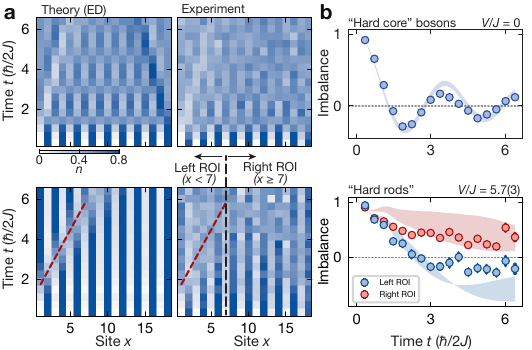}
    \caption{\textbf{Hard rod relaxation dynamics starting from charge-density wave}
    (\textbf{a}) Comparison of the many-body dynamics of a CDW in the hard-core boson (top row; $V/J = 0$) and the ``hard rod" limit (bottom row; $V/J = 5.7(3)$) after a sudden quench of the tunneling energy.
    Here, we illustrate the density evolution calculated using ED ($\bar{n} = 0.5$) and the experimental result with post-selected density $\bar{n} = \si{\numrange[range-phrase = -]{0.44}{0.50}}$.
    For the case with interactions, the red dashed line functions as a guide to the eye and highlights the ballistic spread of an edge defect through the stabilized CDW, leaving a trace by flipping the polarity of the CDW to its left.
    (\textbf{b}) Density imbalance of the system.
    In the hard-core boson limit, we observe the density oscillation of the CDW.
    For hard rods, we observe the stabilization of the CDW on the right indicated by the positive imbalance (red markers; site $\geq 7$), whereas we observe a sign change on the left (blue markers; site $< 7$).
    The shaded areas shows the results obtained by ED for $\bar{n} = \si{\numrange[range-phrase = -]{0.44}{0.50}}$.
    All error bars denote the s.e.m.
    }
    \label{fig:3}
\end{figure}

Proceeding from the two-atom initial state to larger fillings, we use single-site addressing to prepare an initial charge density wave (CDW) with period $\lambda = 2 a_{\mathrm{lat}}$ in the optical lattice and subsequently study its dynamics after a quench.
We first investigate the many-body dynamics in the absence of Rydberg dressing.
For hard-core bosons and fixed boundary walls, the CDW exhibits oscillatory dynamics while maintaining the crystalline-like order, with atoms either occupying the even or the odd sites (see Fig.~\ref{fig:3}a).
We quantify this oscillation by introducing the imbalance defined as
\begin{equation}
    \mathcal{I} = \frac{N_e - N_o}{N_e + N_o}
\end{equation}
where $N_e$ ($N_o$) are the detected atom number on the even (odd) sites.
Comparing our data to ED calculations, including a reduced density $\bar{n} = \si{\numrange[range-phrase = -]{0.44}{0.50}}$, we find excellent agreement (see Fig.~\ref{fig:3}b).
In contrast, operating in the strongly interacting regime $V/J \gg 1$, the hard-core bosons become ``hard rods", featuring an extended exclusion volume and thus inhibiting the occupation of neighboring sites~\cite{naik2024}.
The latter can be explained by the energy gap induced by the bound pair, as discussed in Fig.~\ref{fig:2}a). 
The initially prepared CDW state is thus expected to be stabilized~\cite{guardado-sanchez2021}.
Nevertheless, one can observe defect dynamics within the CDW, if one edge site is unoccupied while the other is occupied, see Fig.~\ref{fig:3}.
In this configuration, the atom closest to the unoccupied left edge site is unimpeded by the extended interactions and can propagate freely through the system.
This process can continue, and the hole defect initially localized on the left edge can ballistically propagate through the system, leaving behind a trace by shifting the entire CDW by one lattice site or, equivalently, flipping the sign of $\mathcal{I}$ in a small subsystem to the left of the defect.
We quantify the different dynamics by evaluating the imbalance on two subsystems, labeled as ``left" (sites $<7$) and ``right" (sites $\geq 7$) region of interest (ROI).
To the right, the imbalance remains positive as the CDW is preserved, whereas on the left the imbalance undergoes a sign change, as the CDW is reordered by the defect motion.
We find qualitative agreement with ED calculations (see Fig.~\ref{fig:3}b) and attribute the remaining deviations to Rydberg decay, imperfect state preparation, and imperfections in the trapping potential, resulting in the loss of individual atoms on the right site.
%


\subsection*{Near-equilibrium density ordering}

\begin{figure}[t!]
    \centering
    \includegraphics[width= 0.5\textwidth]{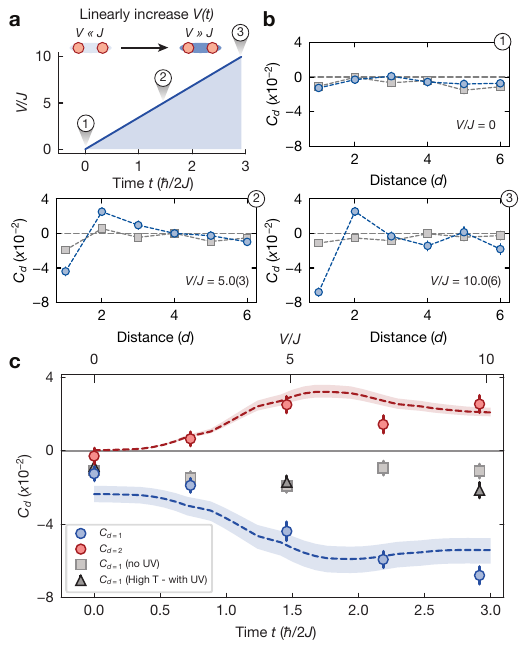}
    \caption{\textbf{Near-equilibrium density-ordering via long-range interactions.}
    (\textbf{a}) Illustration of the experimental protocol.
    Starting from a low-energy ensemble close to half-filling, we linearly increase the nearest-neighbor repulsion up to $ V/J = 10.0(6)$ within $2Jt/\hbar \sim 2.9$.
    Afterwards, we freeze the dynamics and evaluate the density-density correlations.
    (\textbf{b}) Distance-dependent connected correlator $C_d$ for different $V$ and post-selected for mean densities $\bar{n} = \si{\numrange[range-phrase = -]{0.35}{0.52}}$.
    The blue points (gray squares) show the respective correlation strength in presence (absence) of Rydberg-dressed interaction.
    (\textbf{c}) Observed correlations for $d=1$ and $d=2$ during the ramp.
    The dashed lines (shaded areas) represent the theoretical correlations for $k_{\mathrm{B}}T = 1.5J$ ($k_{\mathrm{B}}T = \si{\numrange[range-phrase = -]{1.3}{1.7}}J$) derived by time-dependent MPS.
    The gray squares and black triangles correspond to two reference cases, where we either apply no ramp of the interaction strength or prepare an initial high-energy ensemble before the ramp.
    }
    \label{fig:4}
\end{figure}

Finally, we aim to study the eBHM near equilibrium, where a wealth of interesting quantum phases intimately connected to the long-range interactions are expected~\cite{su2023}.
As a prerequisite, we first realize a 1D low-energy state near the many-body ground state at half-filling in absence of long-range interactions.
We achieve this by preparing a 1D Mott insulator of $9$ atoms on $9$ sites and slowly reducing the lattice depth to enter the itinerant regime.
Subsequently, we adiabatically expand the system into a one-dimensional chain of $17$ sites by lowering the external confinement potential over a timescale of $\sim 60 \hbar/J$~\cite{SI}.
As an upper bound, we estimate the ensemble to have an $\SI{88(2)}{\percent}$ overlap with the initial many-body ground state~\cite{SI}.
Note that a similar overlap has been reported when adiabatically connecting individual atoms, localized in tweezer light, to the lattice ground state~\cite{young2022}. 
After preparing this low-energy ensemble, we next linearly increase the nearest-neighbour interaction from $V=0$ to $V = 10.0(6) J$ over a timescale corresponding to $\sim 2.9$ tunneling times.
Note that matrix-product-state (MPS) simulations~\cite{Schollwoeck2011-DensitymatrixRenormalizationGroup} predict that this short ramp already results in a substantial overlap with the final many-body ground state ($\sim \SI{62}{\percent}$) if the system starts in the many-body ground state before the ramp.
With increasing extended-range interaction strength, the build-up of non-local density-density correlations is expected, as the strong nearest-neighbor repulsion results in crystalline-like structures~\cite{SI}.
To probe this effect experimentally, we analyze the connected correlator $C_d = \langle \hat{n}_i \hat{n}_{i+d}  \rangle - \langle \hat{n}_i \rangle \langle \hat{n}_{i+d}  \rangle$ vs. distance $d$ for various points along the ramp, see Fig.~\ref{fig:4}.
We find that with increasing interaction $V$, the nearest- ($C_{d=1}$) and next-nearest ($C_{d=2}$) neighbor correlation strengths decrease and increase, respectively, indicating the onset of density-ordering in our system.
We observe no significant correlations for distances $d \gtrsim 3$, which is expected for our mean densities of $\bar{n} = \si{\numrange[range-phrase = -]{0.35}{0.52}}$~\cite{SI}.
%
%
Comparing $C_{d=1}$ and $C_{d=2}$ with time-dependent MPS calculations at $k_{\mathrm{B}}T = 1.5J$~\cite{SI}, we find excellent agreement for most interaction strengths (see Fig.~\ref{fig:4}c).
Here, the system's energy is limited by the initial state preparation, whose energy we independently estimate to be $k_{\mathrm{B}}T \sim 2J$~\cite{SI}.
As a reference, when intentionally preparing a system at higher initial energy~\cite{SI}, we find that the correlations vanish (Fig.~\ref{fig:4}c).
Finally, when choosing longer ramp times up to $\sim 8.7$ tunneling times, we observe a slight decrease of the correlation strength, indicating that over longer timescales the Rydberg-state decay and the corresponding heating rate may become relevant~\cite{SI}.

\subsection*{Discussion and conclusion}

In conclusion, our experiments mark the first realization of a Rydberg-dressed extended Bose-Hubbard model with lifetimes long enough to explore a number of intriguing many-body phenomena.
Notably, the interaction strength and range are in a regime compatible with microscopic manipulation of atoms in an optical lattice through a quantum gas microscope, which was essential for all performed experiments.
We anticipate various pathways to improving the coherence time of Rydberg-dressed systems even further, including a cryogenic environment and larger-spaced lattices, which reduce collective losses while allowing to work with longer-lived larger principle quantum numbers, and using alkaline-earth atoms, which feature intrinsically larger transition dipole moments between metastable and Rydberg states, resulting overall in an even more favorable figure-of-merit for limited laser power.
Going forward, the demonstrated unique combination of local microscopic control for initial state preparation, the temporal control over extended-range interactions, and the flexibility in controlling Rydberg-dressed interactions through external fields opens a number of perspectives that are not easily accessible on other platforms.
Our work can, for example, straightforwardly be extended from one-dimensional systems to ladder systems~\cite{kollath2008}, towards two-component mixtures with only one featuring long-range interactions~\cite{li2018}  or to leave the hard-core boson limit and explore the full $V/J-U/J$ phase diagram of the extended Bose-Hubbard model which is expected to include the celebrated Haldane insulator~\cite{dallatorre2006,berg2008}.
Exploring low-energy physics in extended-range interacting fermionic systems, may be relevant to understanding 1D cuprate chains~\cite{chen2021}.
Finally, the constrained dynamics of the hard rod model out-of-equilibrium realized in the presence of the extended interactions is an exciting frontier~\cite{naik2024}, and the arbitrary dynamic control of the interaction strength might open new avenues towards mixed analog-digital quantum simulation approaches~\cite{Morgado2021,malz2023}.
%


\begin{acknowledgments}
We gratefully acknowledge discussions with Simon Hollerith, Waseem Bakr and his entire team, Matthew Eiles, Markus Greiner, Daniel Malz and Markus Heyl.
%
%
%
We acknowledge funding by the Max Planck Society (MPG), the Deutsche Forschungsgemeinschaft (DFG, German Research Foundation) under Germany's Excellence Strategy--EXC-2111--390814868 and through the DFG Research Unit FOR 5522 (project-id 499180199) as well as Project No. BL 574/15-1 within SPP 1929 (GiRyd).
This publication has also received funding under the Horizon Europe program HORIZON-CL4-2022-QUANTUM-02-SGA via the project 101113690 (PASQuanS2.1).
P.W. acknowledges funding through the Walter Benjamin program (DFG project 516136618).
J.Z. acknowledges support from the BMBF through the program ``Quantum technologies - from basic research to market" (Grant No. 13N16265).
K.S. and S.A. acknowledge funding from the International Max Planck Research School (IMPRS) for Quantum Science and Technology.
\end{acknowledgments}



\newpage

\bibliography{bib}
\clearpage

\clearpage
\setcounter{equation}{0}
\setcounter{figure}{0}
\setcounter{table}{0}
\renewcommand{\theequation}{S\arabic{equation}}
\renewcommand{\thefigure}{S\arabic{figure}}
\renewcommand{\thetable}{S\arabic{table}}
\section*{Supplementary Information}

\section{Experimental details}
\label{SM:exp_details}
In this section, we give a detailed description of the initial state preparation, the experimental sequence and the charaterization of the magnetic field gradient.
%


\subsection{Experimental sequence (Fig. 2 and 3)}
\label{SM:exp_sequence}

We start with a near unity-filled 2D atomic Mott insulator (MI) of $^{87}\mathrm{Rb}$ atoms trapped in an optical lattice.
Using single-site addressing~\cite{weitenberg2011,fukuhara2013a}, we prepare three isolated chains with the atoms initialized in the $\ket{F=2, m_F=+2}$ state in the electronic ground state $5S_{1/2}$.
We choose an inter-chain distance of three sites, resulting in negligible inter-chain coupling by the long-range, Rydberg-dressed interaction.
Operating in the atomic limit, we then turn on the magnetic field gradient $\nabla B_y$ along the $y$-direction, suppressing the inter-chain tunneling while tunneling within the chains remains allowed.
Note that with our folded optical lattice~\cite{wei2023}, it is not possible to independently tune the lattice depth in the two tunneling directions via the lattice depth.
Next, using a digital micromirror device (DMD), we project light at $\SI{670}{\nano\meter}$, creating a repulsive potential, which we use to project walls limiting the extent of our 1D chains as well as to compensate on-site potential disorder within the chains.
For all presented experiments studying sudden quench dynamics, we initially reduce the horizontal lattice depth to $\si{\numrange[range-phrase = -]{12}{13}}E_r$, equivalent to tunneling energies of $J/h=\si{\numrange[range-phrase = -]{17}{20}}\si{\hertz}$, in $\SI{0.5}{\milli\second}$.
Here, $E_r = h^2/8ma_\mathrm{lat}^2$ denotes the recoil energy, while $m$ and $a_\mathrm{lat}$ are the atomic mass and the lattice spacing, respectively. 
In order to induce extended-range interactions $V$, we apply an ultraviolet (UV) laser at $\SI{297}{\nano\meter}$ coupling the electronic ground state $5S_{1/2}$ to the Rydberg state $30P_{3/2}$.
After time evolution, we freeze the dynamics by ramping the lattices to $20E_r$ within $\SI{0.3}{\milli\second}$.
%

\subsection{Magnetic field gradient calibration}
\label{SM:magn_gradient}

\begin{figure}[t!]
    \centering
    \includegraphics[width= 0.5\textwidth]{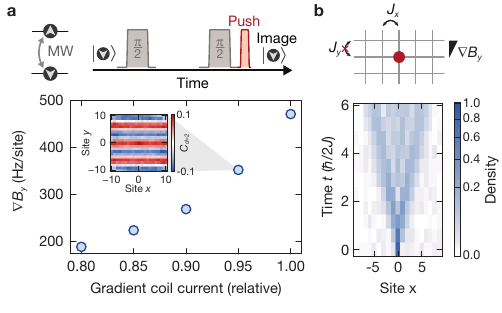}
    \caption{\textbf{Calibrating and benchmarking the magnetic gradient.}
    (\textbf{a}) Ramsey spectroscopy protocol and tunable magnetic gradient.
    We encode our spins in two hyperfine ground states coupled by a microwave drive.
    The magnetic gradient is probed by a Ramsey sequence, where the local spin precession between the two $\pi/2$-pulses results in the build-up of a spin spiral.
    The latter becomes detectable by removing the $\ket{\uparrow}$ states before detection, resulting in a sinusoidal connected correlator $C_{d=2}$, as illustrated in the inset.
    For our experiments we operate at a gradient of $\SI{350(3)}{\hertz}/a_{\mathrm{lat}}$.
    (\textbf{b}) Initializing a single particle within the gradient potential, we observe tunneling within the chain, whereas inter-chain tunneling is negligible.
    }
    \label{fig:S1}
\end{figure}

We characterize the magnitude and the orientation of the magnetic field gradient to isolate the individual chains using Ramsey spectroscopy, see Fig~\ref{fig:S1}a).
To this end, we start by preparing a full 2D MI in the $\ket{\downarrow} = \ket{F=1,m_F=-1}$ hyperfine ground state.
Subsequently, we apply a microwave $\pi/2$-pulse, coupling to the $\ket{\uparrow} = \ket{F=2,m_F=-2}$ state, preparing the atomic ensemble in the superposition state $(\ket{\downarrow}+\ket{\uparrow})/\sqrt{2}$.
For variable holding times of the ensemble, each atom will undergo spin precession depending on the local magnetic field strength.
We map the acquired phase to spin populations by applying a second microwave $\pi/2$-pulse and measure the spins using a resonant push beam that removes the $\ket{\uparrow}$ state before imaging.
If a magnetic gradient is present, the described protocol results in a two-dimensional density wave, as illustrated by the connected correlator in the inset of Fig.~\ref{fig:S1}a).
Fitting a two-dimensional sinusoidal function to these correlators, we derive the orientation and strength of the gradient~\cite{Hild2014}.
For our measurements, we operate with the gradients $\nabla B_y = \SI{350(3)}{\hertz}/a_{\mathrm{lat}}$ and $\nabla B_x = \SI{2(2)}{\hertz}/a_{\mathrm{lat}}$, allowing tunneling along the leg. In contrast, inter-chain tunneling is suppressed, also for potentially occurring higher-order resonant processes (i.e., we ensure $\nabla B_y > U = h \times \SI{225}{\hertz}$).
At this setting, the magnetic field along $\hat{z}$ and $\hat{y}$-direction is $B_z = \SI{4.28}{G}$ and $B_y = \SI{1.31}{G}$, respectively, where $B_x$ is negligible.
We benchmark and ensure 1D tunneling by performing single-particle quantum walks~\cite{weitenberg2011}, as illustrated in Fig~\ref{fig:S1}b).
The latter is further used to calibrate the tunneling energy $J$.
We observe the characteristic light-cone spreading within the chain, whereas tunneling between the chains is negligible.

\subsection{Preparation of low-energy ensemble}
\label{SM:gs_prep}

\begin{figure*}[t!]
    \centering
    \includegraphics[width=\textwidth]{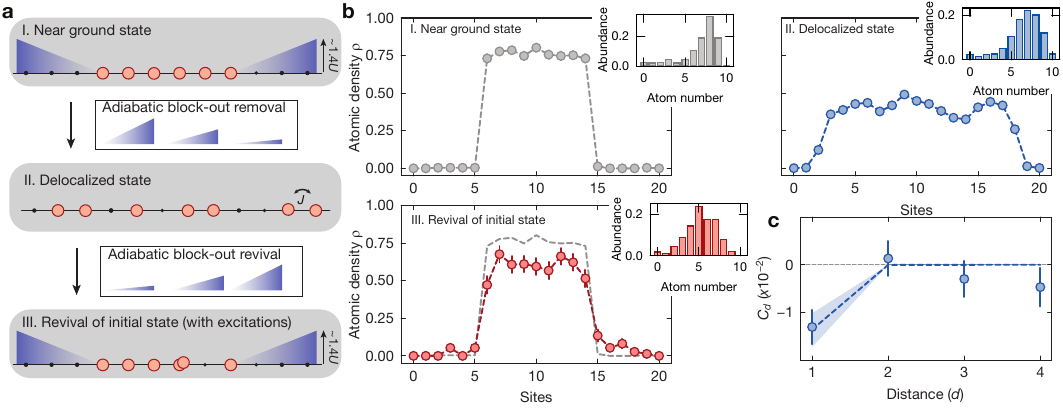}
    \caption{\textbf{Preparation and characterization of the low-energy ensemble.}
    (\textbf{a}) Simplified experimental protocol.
    Starting from a 2D Mott insulator, we deterministically prepare a 1D atomic array using our addressing technique.
    Afterwards, we prepare the remaining ensemble near the many-body ground state by applying a strongly repulsive optical potential to the unoccupied sites.
    We then lower the potential on the blocked-out sites over a timescale of $\sim 60 Jt/\hbar$ to adiabatically connect to a delocalized many-body state.
    Finally, reapplying the repulsive light field and measuring the revival of the density allow us to benchmark the adiabaticity.
    (\textbf{b}) Mean density profiles and atom number distribution (inset) at the three stages of the protocol.
    We start with a uniform density with an average filling of $\SI{76(1)}{\percent}$.
    Lowering the blocked-out sites, the atoms delocalize over the entire system of $17$ sites, while reducing the mean atom number by $0.3(1)$ atoms.
    Ramping up the block-out light once more, we observe the revival of the initial density with an overlap of $\SI{78(2)}{\percent}$.
    For the last case, we further illustrate the initial density as gray-dashed line for reference.
    (\textbf{c}) Distance-dependent connected correlator $C_d$ for the delocalized state post-selected on the fillings $\bar{n} = \si{\numrange[range-phrase = -]{0.35}{0.52}}$. 
    As a guide to the eye we further show the theoretically calculated in-equilibrium connected correlators for $k_{\mathrm{B}}T\sim 2J$ (blue dashed line) and $k_{\mathrm{B}}T\sim \si{\numrange[range-phrase = -]{1.7}{2.3}} J$ (shaded area) obtained using DMRG.
    All error bars denote the s.e.m.
    }
    \label{Sfig:gs_prep}
\end{figure*}

Here, we describe and characterize the preparation of the low-energy ensemble in a single 1D chain, which is then used as a starting point for the interaction ramp shown in Fig.~\ref{fig:4}.
To measure the buildup of density-density correlations near the many-body ground state, the preparation of a sufficiently low-energy state of hard-core bosons at half-filling is required.
To reach such a low-energy state, we first prepare a unity-filled 2D Mott insulator.
Then, we use site-resolved addressing to prepare a single chain of approximately $9$ atoms ~\cite{weitenberg2011, fukuhara2013a}.
Keeping the lattice depth at $60 E_r$, we then adiabatically apply the magnetic gradient and use our DMD to apply a repulsive potential on the empty sites within the target chain.
Note that the repulsive potential has a tapered profile, as illustrated in Fig.~\ref{Sfig:gs_prep}a).
We chose this profile empirically to optimize the adiabaticity of our protocol, as discussed below.
The block-out potential was applied symmetrically on both sides of the system, blocking $4$ sites on each side with $9$ atoms at the center.
Pinning the atoms by the light, we then adiabatically lower the lattice depth to $14.8 E_r$, resulting in a tunneling energy of $J = h \times 11.6(3)$.
We then expand the system into the full chain with $17$ sites by adiabatically reducing the tapered block-out potential.
We achieve this by linearly reducing the light intensity over a timescale of $\sim 60 \hbar / J$.
In the ideal case, the ensemble adiabatically connects to the many-body ground state at a mean filling of $\bar{n} = 0.53$.
We characterize the performance of the preparation by measuring profiles of the mean local atomic density for the preparation steps, as well as the final density after adiabatically ramping up the tapered potential in the end, see Fig.~\ref{Sfig:gs_prep}b) and c).
After the melting sequence, the atoms homogeneously spread over the entire system, featuring small density modulations due to local potential disorder.
Comparing the mean atom number, before and after the melting, we lose $0.3(1)$ atoms on average.
Evaluating the density distribution of the final revived state after adiabatically ramping up the tapered potential, we mostly find the atoms on their original sites.
Comparing the density overlap of the initial with the revived state, we find an overlap of $\SI{78(2)}{\percent}$.
The reduction in atom number by $1.5(2)$ atoms most likely arises due to imperfections in the adiabatic preparation and, consequentially, the formation of doublons, which are removed in our detection scheme due to parity projection~\cite{sherson2010,bakr2009,gross2021a}.
Using the two-way fidelity via the density overlap, we estimate the overlap of the half-filled state with the many-body ground state to be approximately $\SI{88(2)}{\percent}$.
Excitations due to imperfections in the adiabatic ramps lead to finite temperature ensembles.
We can estimate the temperature by comparing our correlators $C_d$ at post-selected densities $\bar{n} = \si{\numrange[range-phrase = -]{0.35}{0.52}}$, with an in-equilibrium DMRG calculation.
We find good agreement for $k_{\mathrm{B}}T \sim 2J$.
Notably, one can derive a similar energy scale by estimating the energy introduced by initial holes.
For our initial filling of $\SI{76(1)}{\percent}$, we expect $2$ holes within the line of $9$ target atoms on average.
Each localized hole carries the energy of $4J$.
Assuming a fully-equilibrated system after the melting protocol, the total energy of $E\sim 8J$ should be equally distributed among the $7$ atoms, meaning each atom carries an energy $E\sim 1.1 J$.
Additional heating might be caused by shape imperfections of the block-out light and a partially diabatic ramp in the melting process.
In the future, we might achieve colder low-energy ensembles by optimizing the initial filling and ramp parameters.
Finally, in Fig.~\ref{fig:4}c) in the main text, we further investigate the time dynamics of a high-energy ensemble as a reference, whose preparation is discussed here.
Similar to the low-energy ensemble, we prepare this state starting with a single chain of approximately $9$ atoms.
Unlike before, we do not apply the repulsive, tapered potential while the remainder of the protocol is kept the same. 
Consequently, when the lattice depth is lowered from $60 E_r$ to $14.8 E_r$, the initial product state cannot adiabatically follow the many-body ground state, resulting in a large energy increase.
The latter becomes evident as we observe no buildup of correlations when $V$ is increased.

\section{Rydberg interactions}
\label{SM:ryd}
In this section, we discuss the advantages and limits of stroboscopic Rydberg dressing and the relevant scalings for the experiment. We present the details of our electronic level scheme, derive the theoretical softcore, and report on an independent calibration of the interaction strength using a spectroscopic Ramsey sequence.

\subsection{Stroboscopic Rydberg dressing: scaling and limits}
\label{SM:strob_ryd_dressing}

\begin{figure}[t!]
    \centering
    \includegraphics[width= 0.5\textwidth]{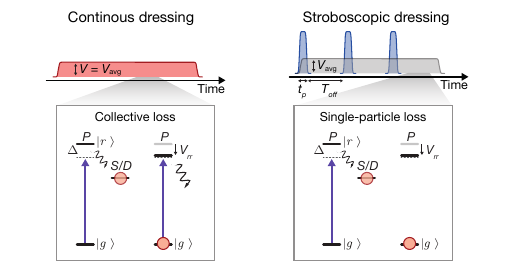}
    \caption{\textbf{Comparison of atom losses by continuous vs. stroboscopic Rydberg dressing.}
    For the continuous dressing (left), off-resonant scattering events followed by a black-body decay populate the neighboring $S$- or $D$-Rydberg manifolds, causing follow-up ``facilitation" losses, as the Rydberg-Rydberg interaction $V_{\mathrm{rr}}$ tunes the coupling laser into resonance.
    For stroboscopic dressing however (right), these many-body losses are mostly suppressed, as the dark-time of the coupling light $T_{\mathrm{off}}$ is significantly larger than the lifetime of the neighboring $S$- or $D$-Rydberg states.
    As the contaminating atoms quickly decay before the next ``on"-time, they cannot cause a secondary loss.
    }
    \label{Sfig:strob_vs_cont_dressing}
\end{figure}

Extending Rydberg dressing towards many-body systems has been mostly limited by the observation of collective avalanche-losses~\cite{goldschmidt2016,zeiher2016,festa2022}.
These are caused by the spurious population of opposite-parity Rydberg states triggered by a black-body decay, which tune the otherwise detuned Rydberg laser into resonance.
The latter can be drastically reduced by applying stroboscopic dressing, which was realized early on~\cite{zeiher2016,Zeiher2017a} and recently confirmed in a different experiment~\cite{hines2023}.
Operating with long dark times, one effectively suppresses the follow-up excitation, turning the collective losses into single-particle losses, as illustrated in Fig.~\ref{Sfig:strob_vs_cont_dressing}.
In addition, stroboscopic dressing has fundamental advantages over continuous dressing due to the effectively larger admixture that can be used.
In the following, we review and present the relevant scalings for stroboscopic Rydberg dressing, following the arguments presented in~\cite{Zeiher2017a}.
For simplicity, we assume an ideal two-level system with states $\ket{g}$ and $\ket{r}$.
Introducing the Rabi rate $\Omega$ and detuning $\Delta$, we can define the admixture $\beta = \Omega / (2\Delta)$.
Operating in the weak dressing regime $\beta \ll 1$, the maximal softcore height amounts to $V_0 = \Omega^4 / (8\Delta^3) = \beta^3 \Omega$ to first order.
Note that in our experiment, we have $V < V_0$ on neighboring lattice sites, as the softcore decreases monotonically with distance.
Fundamentally, dressing is limited by the admixed decoherence rate $\Gamma$, which originates from the single-particle off-resonant scattering rate $\Gamma = \Omega^2/(4\Delta^2) \Gamma_{r} = \beta^2 \Gamma_{r}$, with $\Gamma_{r}$ the finite lifetime of the excited state $\ket{r}$.
Extending this description towards stroboscopically pulsed dressing light, we introduce the duty cycle $D = t_p/(t_p + T_{\mathrm{off}})$, with $t_p$ and $T_{\mathrm{off}}$ the pulse duration and the ``off-time" within the cycle, respectively.
Keeping the $\beta$ constant while reducing $D$ linearly reduces the average stroboscopic softcore height $V_{\mathrm{S,av}} = D V_0$ and the average stroboscopic decoherence rate $\Gamma_{\mathrm{S,av}} = D \Gamma$.
Experimentally, however, we are interested in maintaining the same interaction strength and thus softcore height ($V_{\mathrm{S,av}} = V_0$) while boosting the effective lifetime ($\Gamma_{\mathrm{S,av}} < \Gamma$).
Assuming a fixed Rabi frequency $\Omega$, the interaction strength can be increased by reducing the detuning ($\lvert \Delta_{\mathrm{S}}\rvert < \lvert \Delta \rvert$) such that for a given $D$ one obtains $V_{\mathrm{S,av}} = D\beta_{\mathrm{S}}^3 \Omega =  \beta^3 \Omega = V_0$.
The last condition can be reformulated as $\beta = D^{1/3} \beta_{\mathrm{S}}$ or
\begin{equation}
    \Delta_{S} = D^{1/3} \Delta.
\end{equation}
Operating at the adjusted detuning $\Delta_{\mathrm{S}}$, one naturally extends the lifetime to
\begin{equation}
    \Gamma_{\mathrm{S,av}} = D \beta_{S}^2 \Gamma_{r} = D (D^{-2/3} \beta^2) \Gamma_{r}=D^{1/3} \Gamma.
\end{equation}
A duty cycle of $D = 1/300$ thus leads to an effective lifetime that is increased by a factor of $\simeq 6.7$ while keeping the average interaction strength constant, see Fig~\ref{fig:1}d) in the main text.
The increase in lifetime can be generally understood by comparing the detuning-dependent scalings.
The softcore height scales as $\propto \Delta^{-3}$, whereas the incoherent scattering scales as $\propto \Delta^{-2}$.
As a consequence, the overall figure-of-merit, which is the product of the softcore and the decoherence rate, scales as $\Delta^{-1}$.
Therefore, operating at closer detuning (as in $\lvert \Delta_{\mathrm{S}}\rvert < \lvert \Delta \rvert$), improves the overall lifetime if the averaged softcore height $V_{\mathrm{S,av}}$ is kept constant.
Lowering the detuning is limited by the validity of the weak dressing regime, beyond which the interaction character changes to a many-body interaction.
In addition, for closer detunings, one eventually reaches the Fourier limit of the ``on-pulse" ($\Delta_\mathrm{S} \simeq 1 / t_p $) resulting in the creation of Rydberg excitations.
The latter can be mitigated by increasing $t_p$ and $T_\mathrm{off}$ simultaneously.
Note, however, that in our case the maximal ``off-time" is eventually limited by the external time scales provided by the tunneling energy $J$.
For a stroboscopic extended Hubbard model, the softcore should effectively be considered as a time-averaged quantity, which requires the separation of time scales $1/(t_p + T_{\mathrm{off}}) \ll J$.
For our experiments we typically apply $\gtrsim 25$ pulses during one tunneling timescale.

\subsection{Electronic level scheme and laser parameters}
\label{SM:ryd_level_laser_system}

\begin{figure}[t!]
    \centering
    \includegraphics[width= 0.5\textwidth]{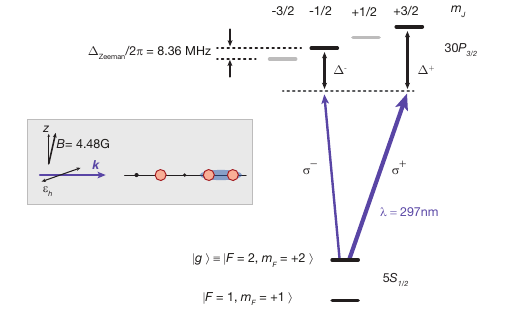}
    \caption{\textbf{Electronic level structure and excitation scheme.}
    The atoms, initially prepared in the $\ket{5S_{1/2}, F = 2, m_{F} = +2}$ state, are off-resonantly coupled to the $\ket{30P_{3/2}}$ Rydberg manifold using a UV laser at $\SI{297}{\nano\meter}$.
    The laser's $\mathbf{k}$-vector is aligned along the chain, whereas the applied magnetic gradient results in a finite offset-field $B = 4.48\,$G.
    Choosing a horizontal polarization $\epsilon_h$ for the UV light, we off-resonantly couple with both the $\sigma^+, \sigma^-$ polarizations.
    }
    \label{Sfig:level_scheme}
\end{figure}

Fig.~\ref{Sfig:level_scheme} shows the relevant electronic states for our atoms and the respective transitions.
Operating with a UV laser at $\SI{297}{\nano\meter}$, we off-resonantly couple the $\ket{5S_{1/2}, F = 2, m_{F} = +2}$ ground state atoms to the $\ket{30P_{3/2}}$ Rydberg manifold.
Here, we operate with a finite offset-field of $B = 4.48\,$G, resulting in a small Zeeman splitting between the $m_J$-levels of $\Delta_{\mathrm{Zeeman}} = \SI{8.36}{\mega\hertz}$.
For our chosen horizontal laser polarization $\epsilon_h$, we couple both to the $\ket{30P_{3/2},m_J = +3/2}$ and the $\ket{30P_{3/2},m_J = -1/2}$ state via the $\sigma^+$-polarization and $\sigma^-$-polarization, respectively.
For our quantum numbers, the ratio between the coupling strengths amounts to $\Omega^+ : \Omega^- \simeq 1 : 0.577$.
Interestingly, despite the closer detuning for the $\sigma^-$ contribution ($\lvert{\Delta_-}\rvert < \lvert{\Delta_+}\rvert$), the nearest-neighbor interaction $V$ is predominantely given by the $\sigma^+$ contribution, due to the stronger coupling ($V \propto \Omega^4$) and the pair potentials at $B = 4.48\,$G.
Note, that we maximize $\Omega^+ \simeq 2\pi \times \SI{20}{\mega\hertz}$ by using a small beam waist of $w_0 \simeq \SI{12}{\micro\meter}$.
In principle, one can further optimize the $\sigma^+$-contribution (and minimize the $\sigma^-$-contribution) by co-aligning the magnetic field with the laser's $\mathbf{k}$-vector.
In this regime however, the softcore would be reduced as demonstrated in Ref.~\cite{zeiher2016}.
Therefore, operating with a vertically aligned field is a compromise.
%

\subsection{Rydberg interactions and softcore estimation}
\label{SM:ryd_inter_softcore_estimation}

\begin{figure*}[t!]
    \centering
    \includegraphics[width=\textwidth]{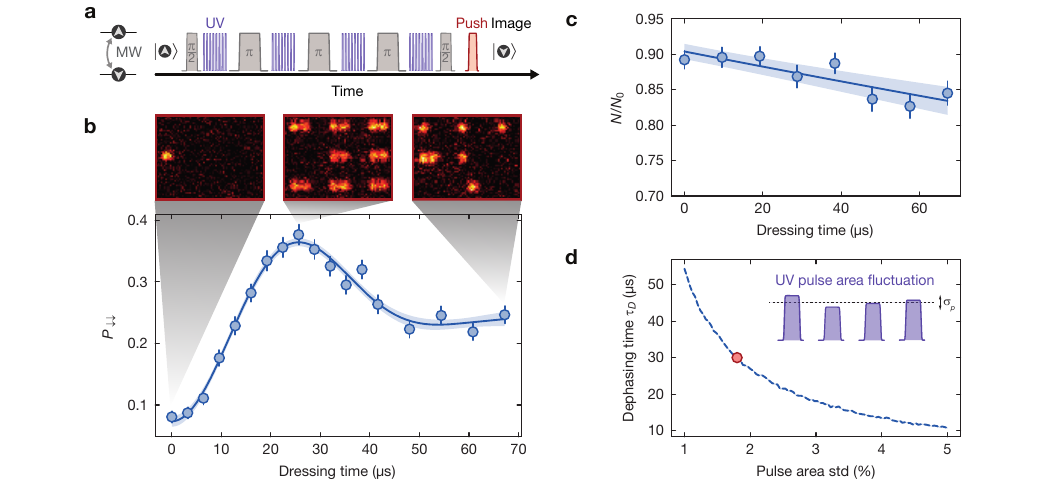}
    \caption{\textbf{Calibration of the nearest-neighbor interaction $V$ using a spectroscopic spin-echo sequence.}
    (\textbf{a}) Spin-echo pulse sequence:
    Starting with spin-polarized atom pairs in the $\ket{\uparrow}$ state, we apply a microwave $\pi/2$-pulse to prepare a coherent superposition of $\ket{\uparrow}$ and $\ket{\downarrow}$.
    We then alternately apply eight UV pulses followed by a microwave $\pi$-pulse, resulting in an interaction-induced phase shift while canceling out the phases acquired from the single-particle light shift.
    For imaging, we apply a final microwave $\pi/2$-pulse and remove all $\ket{\uparrow}$ atoms.
    (\textbf{b}) Pair state probability $P_{\downarrow\downarrow}$ for variable dressing time.
    The solid line and the surrounding shading show the best-fit result with error bars for our empirical model (see Eq.~\ref{seq:damped_pair_oscillation}).
    Insets show exemplary single shots.
    At the beginning, all spins are in the $\ket{\uparrow}$ state and thus removed before imaging.
    At the peak of the oscillation, we observe a large number of flipped pairs.
    At the end, we mostly see individual atoms, as the spins are aligned randomly.
    (\textbf{c}) Single-particle lifetime of the applied echo sequence.
    Fitting an exponential decay (solid line and shaded area), we find a lifetime of \SI{838(213)}{\micro\second}.
    (\textbf{d}) Monte-Carlo simulated dephasing time $\tau_D$ for fluctuating Gaussian noise of the UV pulse area.
    Comparing the simulation with our fitted experimental result (red data point), we estimate an upper bound for the UV pulse area fluctuation with a standard deviation of $\sigma_p \lesssim \SI{1.8}{\percent}$.
    All error bars denote the s.e.m.
    }
    \label{Sfig:softcore_calib}
\end{figure*}

We calculate the Rydberg pair potentials using the software ``pairinteraction"~\cite{weber2017}. 
First, we calculate the distance-dependent interaction shift $V_{\mathrm{rr}}(d)$ between two Rydberg atoms in the $\ket{r} = \ket{30\mathrm{P}_{\mathrm{3/2}}, m_{j} = +3/2}$ state at a distance $d$.
The relevant states and their respective relative coupling to the ground states are illustrated in the inset of Fig.~\ref{fig:1}a) in the main text.
Next, we derive the softcore height and shape by diagonalizing the two-atom Hamiltonian 
\begin{equation}
\begin{split}
H(d) =& \frac{\hbar}{2} \Omega(\ket{rg}\bra{gg}+\ket{gr}\bra{gg}) - \hbar \Delta(\ket{rg}\bra{rg}+\ket{gr}\bra{gr}) \\ 
&+ \frac{\hbar}{2}\sum_{r',r''} \Omega C_{r'r''} (\ket{r'r''}\bra{rg} + \ket{r'r''}\bra{gr}) \\
&+ \frac{\hbar}{2\pi}\sum_{r',r''}(-2\Delta + V_{\mathrm{rr}}^{r'r''}(d))\ket{r'r''}\bra{r'r''}.
\end{split}
\end{equation}
Here, $g$ is the electronic ground state, $\ket{r'r''}$ are the Rydberg pair states and $C_{r'r''} = \braket{r'r''}{rr}$ is the overlap of $\ket{r'r''}$ with the bare state $\ket{rr}$.
We account for all Rydberg states with $\lvert V_{\mathrm{rr}}^{r'r''} \lvert<\SI{3}{\giga\hertz}$.
For our paramters of $\Omega = 2\pi\times \SI{20}{\mega\hertz}$, $\Delta = 2\pi\times \SI{-60}{\mega\hertz}$ and $D = 1/600$, we obtain $V/h = \SI{94}{\hertz}$ for the nearest-neighbour interaction.
Note that the latter is calculated for the peak intensity of our coupling laser.
In our experiments, we intentionally displace the center of the Gaussian beam by about three lattice sites to ensure that the AC stark shift does not compensate for the magnetic field gradient due to beam pointing drifts.
Given our beam waist of $w_0 \simeq \SI{12}{\micro\meter}$ and the fact that we are averaging our results over several chains separated by three rows, we estimate an average nearest-neighbor interaction on the order of $V/h \simeq \SI{80}{\hertz}$.
So far, all estimations only include the $\sigma^{+}$ contribution.
Note, however, that we estimate the maximal softcore contribution for the $\sigma^{-}$ contribution to amount to $V(\sigma^{-})/h \lesssim \SI{9}{\hertz}$.
%

%

\subsection{Softcore calibration by spin echo sequence}
\label{SM:softcore_calib_spin_echo}

Similar to previous studies~\cite{zeiher2016,guardado-sanchez2021,steinert2023,holland2023}, we calibrate the nearest neighbor interaction $V$ using a spectroscopic spin-echo sequence, as illustrated in Fig.~\ref{Sfig:softcore_calib}a).
Here, we operate with two hyperfine ground states and make use of the fact that the UV beam only couples to the upper state $\ket{\uparrow} = \ket{F = 2, m_F = +2}$, while the lower manifold $\ket{\downarrow} = \ket{F = 1, m_F = +1}$ remains unaffected.
During the echo sequence, the atoms experience phase shifts by both the single-particle light shift $\delta$ and the two-particle interaction shift $V$.
Ideally, the spin-echo sequence removes the contributions of $\delta$, providing direct access to $V$.

In our experiment, we prepare isolated pairs to derive a clean $V$ while suppressing the beating with other frequencies~\cite{zeiher2017}.
For the spin-echo sequence, we start in the $\ket{\uparrow}$ state and first drive a microwave $\pi/2$-pulse, preparing a coherent superposition between $\ket{\uparrow}$ and $\ket{\downarrow}$.
We then apply eight UV pulses before driving a microwave $\pi$-pulse.
Note that we apply a triple spin-echo sequence with a total of three $\pi$-pulses, improving the dephasing time $\tau_D$.
At the end of the sequence, we apply another $\pi/2$-pulse.
In absence of interactions, all spins return to the original $\ket{\uparrow}$ state, which is pushed out before imaging the $\ket{\downarrow}$ state.

We derive $V$ by monitoring the coherent pair flips between the pair states $\ket{\uparrow\uparrow} \leftrightarrow \ket{\downarrow\downarrow}$.
For an ideal spin-echo sequence starting in $\ket{\uparrow\uparrow}$, the flipped-pair probability is given by
\begin{equation}
    P_{\downarrow\downarrow} = \frac{1}{2} \left(1 - \mathrm{cos}\left(\frac{Vt}{2\hbar}\right) \right),    
\end{equation}
as derived in the references~\cite{zeiher2016,holland2023,bao2023}.
In presence of decoherence or dephasing, the spins will eventually align randomly, resulting in the damping of the oscillation.
Motivated by the recent observation of damped pair oscillations between dipolar molecules in tweezers~\cite{holland2023}, we make use of a similar phenomenological model to derive $V$.
The model is described by a doubly damped oscillation, capturing both particle loss $\tau_{\mathrm{L}}$ and dephasing $\tau_{\mathrm{D}}$ due to Gaussian noise in the UV pulse area:
\begin{equation}
 P_{\downarrow\downarrow} = A e^{-t/\tau_{\mathrm{L}}} \left( 1 - e^{-t^2/(2 \tau_{\mathrm{D}}^2)} \cdot \mathrm{cos} \left(\frac{Vt}{2\hbar}\right) \right)  
 \label{seq:damped_pair_oscillation}
\end{equation}

\begin{figure*}[t!]
    \centering
    \includegraphics[width=\textwidth]{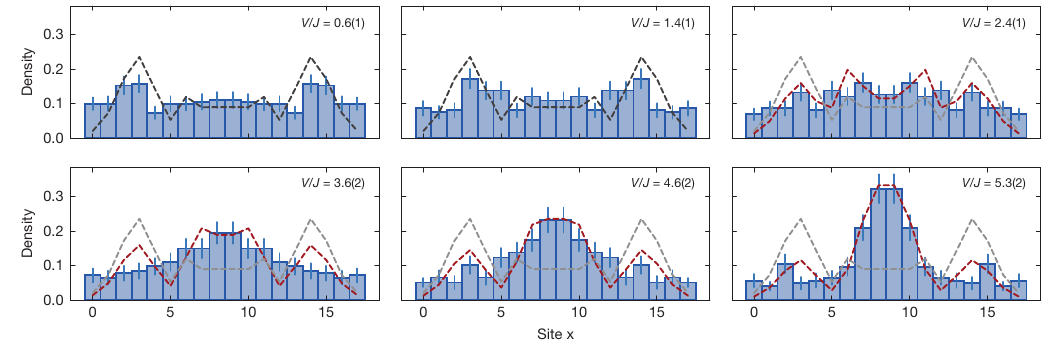}
    \caption{\textbf{Fitting the group velocity of the repulsively-bound pair state.}
    Operating at a fixed evolution time of $2Jt/\hbar \simeq 7$, we step the nearest-neighbor interaction $V$ and fit the group velocity of the pair.
    The group velocity of the unbound atoms are kept constant for the fit and are independently derived from case \textcircled{5} of the main text.
    For $V > V_c$, we can fit a double light-cone, illustrated by the red dashed lines.
    For lower values ($V < V_c$), our data becomes consistent with an ED calculation assuming the absence of bound pairs, shown by the black, dashed lines.
    The latter is further illustrated as gray, dashed lines in the other cases to demonstrate the striking difference arising from the presence of the bound state.
    }
    \label{sfig:step_softcore_group_velocity}
\end{figure*}

Fig.~\ref{Sfig:softcore_calib}b) shows our experimental results and the best fit result of the model, from which we obtain $V/h = \SI{34.1(13)}{\kilo\hertz}$ and $\tau_D = \SI{30(1)}{\micro\second}$.
Operating with a duty cycle of $D = 1 / 600$, we obtain a nearest-neighbor interaction of $V/h = \SI{57(2)}{\hertz}$.
Note that this value is on the order of our theoretical estimate $V/h \simeq \SI{80}{\hertz}$ derived before.
%
%
Interestingly, according to the fit, the time-scale related to particle loss $\tau_{\mathrm{L}}$ is negligible and thus not responsible for the observed damping.
The latter is consistent with the measured atomic lifetime of $\SI{838(213)}{\micro\second}$, see Fig.~\ref{Sfig:softcore_calib}c).
Shot-to-shot fluctuations of the UV pulse area can quickly result in dephasing, as even fluctuations on the few-percent level result in strong variations of induced phase shift $\phi = \delta \cdot t$~\cite{zeiher2017}.
The difference in the accumulated phase shifts between the various $\pi$-pulses will eventually dominate, resulting in randomly-aligned spins in the end.
We can simulate this phenomenon and derive a dephasing time $\tau_{\mathrm{D}}$ using a classical Monte-Carlo simulation.
To this end, we model the rotations on the Bloch sphere including the microwave pulses as well as the $32$ UV pulses.
In our model we assume the pulse area of the UV pulses to follow a Gaussian distribution with standard deviation $\sigma_P$.
For each $\sigma_P$, we can derive a dephasing time $\tau_D$, illustrated in Fig.~\ref{Sfig:softcore_calib}d).
Assuming that our observed time-scales originate exclusively from this effect, we can derive an upper bound for the relative pulse area fluctuations of $\sigma_p \lesssim \SI{1.8}{\percent}$.
%

\section{Fit functions and extended data sets}
\label{SM:ext_data}

In this section, we discuss the fit function utilized for the repulsively-bound pair states and provide additional data sets including three-particle bound states and longer ramps for the near-equilibrium dynamics.

\subsection{Double light-cone fit function}
\label{SM:fit_functions}

The probability of finding a single atom undergoing a quantum walk in an optical lattice can be described by a Bessel function of the first kind
\begin{equation}
    \rho_i(t) = \lvert \mathcal{J}_i(2Jt) \rvert ^2,
\end{equation}
with $i$ denoting the lattice sites~\cite{preiss2015}.
The coherent interference between all pathways results in ballistic transport with an interference pattern in the center and a clear wavefront propagating with group velocity $2Jt/\hbar$.
Initializing a pair state, i.e., two atoms on neighboring sites, while operating in the hard-core boson limit without Rydberg dressing, the atomic density is simply defined by the incoherent sum of two Bessel functions
\begin{equation}
    \rho_i(t) = (\lvert \mathcal{J}_i(2Jt) \rvert ^2 + \lvert \mathcal{J}_{i+1}(2Jt) \rvert ^2)/2.
    \label{Seqn:two_atoms}
\end{equation}
Here, the density still spreads ballistically with $2Jt/\hbar$, but the interference pattern in the center vanishes, as seen in our experimental data in Fig.~\ref{fig:2}b) (case \textcircled{1} and \textcircled{2}).
Extending the model to include the bound pair state is straightforward as the bound state also expands ballistically, following Eq.~\ref{Seqn:two_atoms} with a reduced tunneling energy $J_b$ compared with free pair tunneling energy $J_f$.
To fit the model to our data, we further incorporate the relative amplitude of the free pair $A_f$ and bound state $A_b$ (with $A_f + A_b = 1$), respectively, resulting in
\begin{align}
        \rho_i(t) & = A_f \cdot (\lvert \mathcal{J}_i(2J_f t) \rvert ^2 + \lvert \mathcal{J}_{i+1}(2J_f t) \rvert ^2)/2 \notag \\
        & + A_b \cdot (\lvert \mathcal{J}_i(2J_b t) \rvert ^2 + \lvert \mathcal{J}_{i+1}(2J_b t) \rvert ^2)/2.
        \label{Seqn:quantum_walk_pair}
\end{align}
%

\subsection{Repulsively-bound pair states}
\label{SM:group_velocity}

In Fig.~\ref{fig:2} in the main text, we characterize the quench dynamics of repulsively-bound pair states.
We investigate the full-time evolution for the presented cases \textcircled{1}-\textcircled{5}, allowing us to fit the inner light-cone as discussed above.
In Fig.~\ref{fig:2}c) we compare these fit results with the exact solution derived in Ref.~\cite{valiente2009}.
There, we also present the results of additional measurements, whose evaluation is discussed here.
Unlike the measurements in the main text, where we step the evolution time for a fixed interaction strength $V/J$, the additional data sets are obtained for a fixed tunneling time $2Jt/\hbar \simeq 7$ while the nearest-neighbor interaction is stepped.
These measurements were taken at the same time as case \textcircled{5} of the main text, such that we can use its free light-cone as a fixed quantity while deriving the group velocity of the bound state.
Fig.~\ref{sfig:step_softcore_group_velocity} shows the experimental results with the corresponding best fits using Eq.~\ref{Seqn:quantum_walk_pair}.
For interactions below the critical value ($V < V_c$), our measurements are consistent with an ED calculation assuming the absence of bound states (see black, dashed line).
For larger values ($V > V_c$), we observe a clear deviation and the emergence of a second, inner light-cone.
The best-fit result is shown by the red dashed line.

\subsection{Three-atom bound state}
\label{SM:three_particle_bound_state}

\begin{figure}[t!]
    \centering
    \includegraphics[width= 0.5\textwidth]{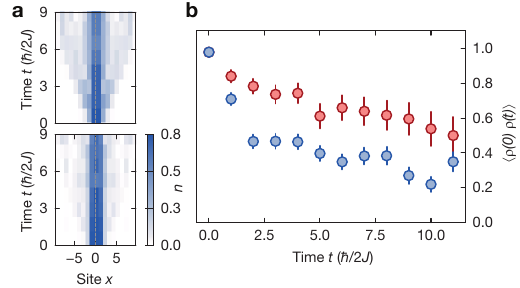}
    \caption{\textbf{Repulsively-bound dimers and trimers.}
    (\textbf{a}) Symmetrized two-particle and three-particle quench dynamics for $V/J = 4.9(2)$ in the upper and lower part, respectively.
    The vertical gray dashed lines show the symmetrization axis.
    (\textbf{b}) Atomic density at the two, three central sites for the two-atom (blue), three-atom (red) case.
    The three-atom bound state experiences an effective larger repulsion and has a lower group velocity, resulting in a larger density at the center over time.
    }
    \label{Sfig:2_vs_3_atoms}
\end{figure}

The presented experiments can be naturally extended towards larger many-body bound states.
The dynamics of these longer strings, often labeled as ``$l$-bits" or ``$l$-strings" of length $l$, have been extensively studied theoretically~\cite{ganahl2012b,aleiner2021} and have recently been observed experimentally with microwave photons~\cite{morvan2022}.
Using a Bethe ansatz, one can derive the exact bandstructure of these larger many-body bound states~\cite{aleiner2021}.
Interestingly, their resulting tunneling dynamics is exponentially suppressed $\propto e^{-l}$, which implies that larger complexes are effectively frozen.
In our setup, we can probe the tunneling dynamics of larger ``$l$-strings" by preparing $l$ neighboring atoms in a row.
Here, we compare the three-atom case ($l=3$) with the two-atom case from the main text.
Here, we operate with $V/J = 4.9(2)$ for both cases, giving us direct insights into the influence of $l$.
Fig.~\ref{Sfig:2_vs_3_atoms}a) shows the symmetrized density-evolution for both cases.
For $l=2$, we see a clear double light-cone structure featuring the single-particle light-cone and a slowly spreading central light-cone of the bound pair.
For $l=3$, we have a strong density peak at the center, while the single-particle light-cone is partially visible but strongly suppressed, consistent to previous observations~\cite{morvan2022}.
Given our time-scales of $2Jt/\hbar \lesssim 11$, we are unable to detect any higher-order tunneling for the three-atom bound state.
Nevertheless, we can probe for its robustness by evaluating the time evolution of the mean atomic density on the original sites $\langle \rho(0) \rho(t) \rangle$, as illustrated in Fig.~\ref{Sfig:2_vs_3_atoms}b).
As expected, we observe the three-atom case to be strongly localized on its original sites.
%

%
\subsection{Near-equilibrium density ordering: longer ramps}
\label{SM:longer_ramps}

A requirement for studying adiabatic preparation in the eBHM near equilibrium is to tune the nearest-neighbor interaction slowly compared to the tunneling timescale $\hbar/2J$.
Fig.~\ref{fig:4} of the main text demonstrates a first step along these lines, where we tune $V$ over the course of $\sim 2.9$ tunneling time scales.
In this section, we present additional data for longer ramp durations of $\sim 5.8$ and $\sim 8.7$ tunneling time scales, respectively.
In the absence of any imperfection such as Rydberg-losses and heating, we would expect the slower ramps to result in a larger overlap with the many-body ground state, featuring stronger nearest-neighbour ($C_{d=1}$) and next-nearest neighbour correlations ($C_{d=2}$).

\begin{figure}[t!]
    \centering
    \includegraphics[width= 0.5\textwidth]{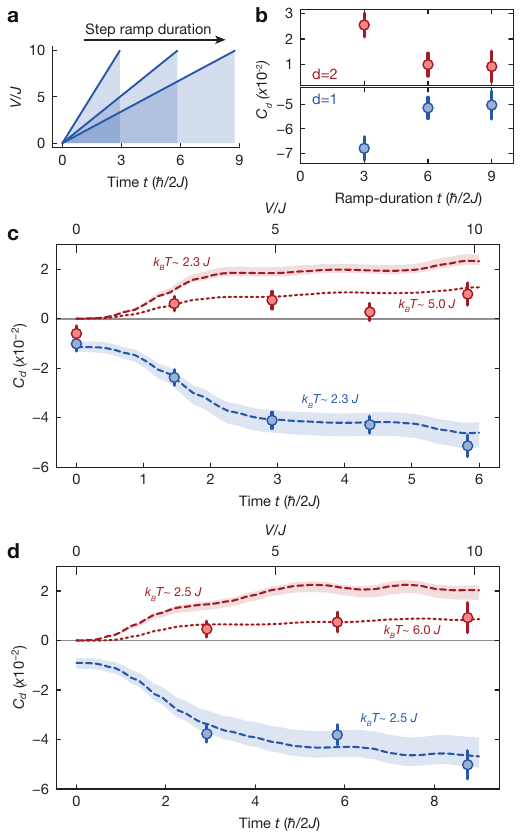}
    \caption{\textbf{Dynamically-induced density-ordering for longer ramp durations.}
    (\textbf{a}) Illustration of the experimental protocol.
    Starting from a low-temperature ensemble close to half-filling, we linearly increase the nearest-neighbor repulsion up to $ V/J = 10.0(6)$.
    Here, we probe three different ramp durations $2Jt\hbar \sim \{2.9, 5.8, 9.7\}$.
    (\textbf{b}) Final correlations for $d=1$ and $d=2$ at the end of the different ramps.
    (\textbf{c}) Build-up of correlations for $d=1$ and $d=2$ during the ramp.
    The dashed lines (shaded areas) represent the theoretical correlations for $k_{\mathrm{B}}T = 2.3J$ ($k_{\mathrm{B}}T = \si{\numrange[range-phrase = -]{2.0}{2.5}}J$) derived by DMRG.
    (\textbf{d}) Similar to (c), yet for the longest ramp duration.
    The dashed lines (shaded areas) represent the theoretical correlations for $k_{\mathrm{B}}T = 2.5J$ ($k_{\mathrm{B}}T = \si{\numrange[range-phrase = -]{2.3}{3.0}}J$) derived by DMRG.
    All measurements are post-selected for mean densities $\bar{n} = \si{\numrange[range-phrase = -]{0.35}{0.52}}$
    }
    \label{Sfig:longer_ramps}
\end{figure}

Fig.~\ref{Sfig:longer_ramps} summarizes our observed correlators for the various ramp durations.
In Fig.~\ref{Sfig:longer_ramps}b) we first compare the correlators at the end of each ramp.
Interestingly, the correlators are the strongest for the shortest ramp and are slightly reduced for the longer ramps.
The latter might be explained by an overall energy increase of the ensemble caused by Rydberg losses.
Note that while for the shortest ramp we lose on average $\sim 0.9$ atoms, we lose up to $\sim 2.0$ atoms for the longer ramps.
To quantify the resulting heating effect, we compare the experimental time evolution with dynamical DMRG calculations in Fig.~\ref{Sfig:longer_ramps}c) and d).
Here, we make two key observations.
First, evaluating $C_{d=1}$, we observe a monotonic increase of the ensemble temperature to $k_{\mathrm{B}}T = 2.5J$.
Second, for longer ramp durations, we observe a temperature mismatch between $C_{d=1}$ and $C_{d=2}$, indicating that the system did not fully re-thermalize after an atom-loss event.
In the future, there are different pathways to explore these processes in more detail.
Experimentally, it would be interesting to probe these ramps with lower, final nearest-neighbor interaction $V_f$.
Operating at $V_f / J > 2$ and at half-filling one would expect the onset of density-ordering, while the reduced Rydberg-losses allow to probe longer ramp times where a cleaner study of the microscopic processes is feasible.
On the theoretical side, it would be interesting to extend our current DMRG model towards an open quantum system, including Rydberg losses.
%

\section{Numerical methods}

\subsection{DMRG and time evolution}

\begin{figure}[t!]
    \centering
    \includegraphics[width= 0.5\textwidth]{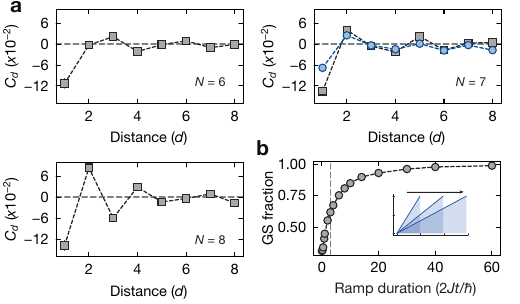}
    \caption{\textbf{Ground state properties.}
    (\textbf{a}) Distance-dependent connected correlator for the many-body ground state at $V/J = 10.4$ and variable atom number $N$.
    The experimental data is illustrated by the blue data points in the second panel.
    Experimentally, we observed a mean atom number of $\langle N \rangle = 6.9$.
    (\textbf{b}) Ground state fraction after applying a linear ramp of variable duration.
    The evolution starts from the perfect many-body ground state at $V = 0$.
    The vertically dashed line illustrates the ramp duration of Fig.~\ref{fig:4} of the main text.
    }
    \label{Sfig:ground_states}
\end{figure}

For the numerical reference data presented in Fig.~\ref{fig:4} and Fig.~\ref{Sfig:gs_prep} we use the matrix-product-state (MPS) formalism~\cite{Schollwoeck2011-DensitymatrixRenormalizationGroup} to simulate the one-dimensional extended Bose-Hubbard model
\begin{align}
\begin{split}
    \hat{H}_\mathrm{eBHM} = \
    & - J \sum_{i} (\hat{a}^\dagger_{i} \hat{a}_{i+1} + \mathrm{h.c.})
    + \frac{U}{2} \sum_{i} \hat{n}_{i} (\hat{n}_{i} - 1) \\
    & + V \sum_{i} \hat{n}_{i} \hat{n}_{i+1}
    + \mu \sum_{i} \hat{n}_{i} \ ,
\end{split}
\end{align}
with the lattice size $L=17$ realized in the experiment.
Here, $\hat{a}^{(\dagger)}_{i}$ is the bosonic annihilation (creation) operator at lattice site $i$ and $\hat{n}_i = \hat{a}^\dagger_{i} \hat{a}_{i}$ is the particle number operator.
Due to the large on-site interaction $U/J \gg 1$, we model the system using hard-core bosons, i.e. $n_i \in \{ 0, 1 \}$.
Consequently, the model is characterized by the tunneling energy $J$ and nearest-neighbor repulsion $V$. 
We model finite temperatures and particle number $\langle \hat{N} \rangle$ slightly below half-filling by introducing a chemical potential $\mu = 0.6 \: J$.
For the ground-state calculations, we employ the density-matrix renormalization group (DMRG)~\cite{White1992-DensityMatrixFormulationb, White1993-DensitymatrixAlgorithmsQuantuma} algorithm and exploit the $U(1)$ symmetry corresponding to particle-number conservation to work in a fixed $N$-sector.
Finite-temperature calculations are performed via imaginary-time evolution starting from an infinite-temperature state, where purification~\cite{Feiguin2005-FinitetemperatureDensityMatrix} is used to describe mixed states using MPS.
Up to an inverse temperature $\beta = 1 / (k_\mathrm{B} T) = 0.2 \: J$, we use the global Krylov method~\cite{Garcia-Ripoll2006-TimeEvolutionMatrixa} to ensure accurate imaginary-time evolution at low bond dimensions. Following this threshold, the time-dependent variational principle~\cite{Haegeman2016-UnifyingTimeEvolutiona} is used in its two-site version (2TDVP) with step size $\Delta \beta = 0.02 \: J$. \\
To simulate the system's dynamics, real-time evolution is performed using 2TDVP with step size $\Delta t = 0.025 \: \hbar / J$, chosen to be small to allow keeping the time-dependent Hamiltonian constant during each step. All calculations use a maximum bond dimension of $m = 256$. \\
Since we model a closed quantum system, the particle number $\langle \hat{N} \rangle$ stays constant throughout the simulated dynamics, whereas in the experiment $\langle \hat{N} \rangle$ typically drops by $\Delta \langle \hat{N} \rangle \sim 0.5$ during ramps due to Rydberg losses.
To ensure the accuracy of our comparison between simulation and experiment, we explicitly check the impact that discrepancies in particle number have on the nearest-neighbor (NN) and next-nearest-neighbor (NNN) correlations at the elevated temperatures $1 \lesssim k_\mathrm{B} T / J \lesssim 10$ considered in this work.
Lowering the chemical potential to $\mu = 0.5 \: J$ compared to $\mu = 0.6 \: J$ used for all comparisons induces a shift of
\begin{align}
\begin{split}
    \Delta \langle \hat{N} \rangle
    &= \langle \hat{N} \rangle(\mu = 0.5 \: J) - \langle\hat{N} \rangle(\mu = 0.6 \: J) \\
    &= 7.31 - 7.03 = 0.28
\end{split}
\end{align}
in the $k_\mathrm{B} T = 1.5 \: J$ reference shown in Fig.~\ref{fig:4}c). Throughout the simulated ramp, this leads to a maximum deviation of the NN correlations by $1.3 \%$ and of the NNN correlations by $4.6 \%$ relative to the correlator's final values at the end of the ramp.
In comparison, changing the temperature by $\Delta (k_\mathrm{B} T) = \pm 0.2 \: J$ as indicated by the shaded areas in Fig.~\ref{fig:4}c) corresponds to a maximum deviation of $14.2 \%$ and $18.5 \%$ of the NN and NNN correlations, respectively.
Consequently, we expect discrepancies caused by deviations in $\langle \hat{N} \rangle$ to lie within the shaded areas provided for different temperature ranges. \\
On the flip side, the comparison also indicates that the longer-range NNN correlations are significantly more susceptible to changes in the particle number than the nearest-neighbor correlations -- even at elevated temperatures. In the following section, we relate this feature to the distinct density-ordering patterns of the system's ground states at different $N$. \\

\subsection{Ground state properties}

%
Finally, we use our MPS model to derive the many-body ground state for variable atom number $N$.
To study the charge order, we consider the connected density-density correlation function
\begin{align}
    C_d = \langle \hat{n}_i \hat{n}_{i+d}  \rangle - \langle \hat{n}_i \rangle \langle \hat{n}_{i+d}  \rangle
\end{align}
and compare the ground-state properties to experimental data in Fig.~\ref{Sfig:ground_states}. Interestingly, the data ($\langle \hat{N} \rangle = 6.9$) matches the correlation pattern of the $N = 7$ ($\bar{n} = 0.41$) ground state well. This pattern is clearly distinct from those of the $N = 6$ and $N = 8$ ground states, suggesting that, despite the relatively short ramp duration and elevated temperature, density ordering is established on the many-body level in the experimental system. 
In addition, considering the system's ground states allows us to estimate the time needed to adiabatically prepare the density-ordered states with high fidelity.
Starting from the many-body ground state at $V = 0 \: J$, real-time evolution is performed for linear ramps to $V_\mathrm{f} = 10.4 \: J$ with a range of final times $0.2 \: \hbar / 2 J \leq t_\mathrm{f} \leq 60 \: \hbar / 2 J$.
We find that the short ramping protocol with $t_\mathrm{f} = 3.0 \: \hbar / 2 J$, akin to the one presented in Fig.~\ref{fig:4}, already results in a simulated preparation fidelity of $F = 62 \%$. The longer preparation protocols discussed in Section~\ref{SM:longer_ramps} allow preparation fidelities up to $90 \%$ in principle but are limited by heating.


\end{document}